\documentclass[twocolumn,showpacs,preprintnumbers,amsmath,amssymb]{revtex4}
\usepackage{graphicx}
\usepackage{dcolumn}
\usepackage{bm}
\usepackage{mathrsfs}
\begin{document}

\title{Optical Analog of the Iordanskii Force in a
           Bose-Einstein Condensate }
\author{U.\ Leonhardt and P.\ \"Ohberg}
\address{School of Physics and Astronomy, University of St Andrews,
North Haugh, St Andrews, Fife, KY16 9SS, Scotland} 
\begin{abstract}
A vortex in a Bose-Einstein condensate generates the optical
analog of the Aharonov-Bohm effect when illuminated with slow
light. In contrast to the original Aharonov-Bohm effect the
vortex will exchange forces with the light that lead to a
measurable motion of the vortex.
\end{abstract}
\date{today}

\maketitle

\section{Introduction}

Bose-Einstein condensates of alkali atoms \cite{BEC} are
fascinating superfluids \cite{Tilley}, combining conceptual
simplicity with a wide experimental repertoire of optical
manipulation and imaging. For example, vortices \cite{Vortices}
have been made in optical analogs of the classic rotating-bucket
experiment \cite{Tilley}. The interference of a vortex with a
homogeneous condensate \cite{BWTD} has revealed for the first
time \cite{Inouye} the phase structure of quantum vortices
\cite{Tilley}. Quantum shock waves, shedding topological defects,
have been launched using ultra-compressed pulses of slow light
\cite{Dutton}. Here we suggest that the optical Aharonov-Bohm
effect \cite{LP} of slow light in Bose-Einstein condensates
\cite{Hau} can, quite literally, shed light onto a controversial
subject in superfluidity.

The problem arises in analogs
\cite{SAB,OAB,Other} of the Aharonov-Bohm effect \cite{AB}. 
In Aharonov and Bohm's original scenario \cite{AB} a thin solenoid
generates a magnetic field inside yet no field outside the coil
where, however, the magnetic vector potential forms a vortex. A
charged particle would pass the magnetic vortex along a straight
line, because no force is acting on it, but a charged wave forms
a characteristic interference pattern \cite{AB}. The phase $S$ of
the wave function $\psi$ compensates for the vortex of the vector
potential ${\bf A}$ according to the relation \cite{Tilley}
\begin{equation}
m{\bf v} = \hbar \nabla S + q {\bf A}
\end{equation}
for a particle with mass $m$, charge $q$ and velocity ${\bf v}$.
Within the accuracy of a semiclassical description, the
Aharonov-Bohm interaction does not change the momentum density
$m|\psi|^2 {\bf v}$ of the wave. Consequently, according to the
principle of {\it actio et reactio} we would not expect that the
wave exerts any force back onto the magnetic vortex \cite{Shelankov}.
 
However, this argument does not necessarily apply to analogs
\cite{SAB,OAB,Other} of the Aharonov-Bohm effect.
Waves in fluids, such as sound \cite{SAB} or light \cite{OAB} 
experience a flow {\bf u} as an effective vector potential {\bf A},
as long as the flow speed $u$ is much smaller than the group
velocity of the wave. A fluid vortex generates the equivalent
of the Aharonov-Bohm effect \cite{SAB,OAB}.
On the other hand, for waves in fluids, the gradient of the phase
represents the true momentum, whereas the equivalent of 
$m{\bf v}$ plays the role of a pseudomomentum \cite{Stone1}.
Consequently, the Aharonov-Bohm phase slip 
leads to a momentum transfer between the wave and the fluid,
and hence to a force that sets the vortex into motion. 
The moving vortex is then subject to another force, the
Magnus force of the fluid \cite{HallVinen,Kopnin}.
As we show in this paper, 
the two forces combined generate an interesting dynamics,
where the vortex spirals towards the incident wave. 

In superfluids such as $^4\mbox{He}$ \cite{Tilley}, 
the superfluid component 
is made of a Bose-Einstein condensate and 
the normal component consists of elementary excitations, 
i.e. sound waves in the large-wavelength limit.
Suppose that there is no net flow of the normal component.
In this case the excitations are not directed in any way,
similar to the molecules of a gas. The combined effect of
the individual forces between each excitation and the vortex 
causes friction \cite{Kopnin,Volovik}, 
as has been first calculated by Iordanskii \cite{Iordanskii}.
From the very beginning \cite{Sonin}
Iordanskii's force has been subject to bitter controversy
\cite{Sonin,Controversy,Stone2}. Experiments \cite{Friction}
indicate that the force does indeed contribute substantially to the mutual
friction between the superfluid and the normal component, yet
there is no direct experimental evidence for the Iordanskii force
as a non-dissipative mechanical force on vortices. Here we point
out that the optical equivalent of the Iordanskii force can be
seen in action in the Aharonov-Bohm effect \cite{LP} of slow light
\cite{Hau} at vortices \cite{Vortices} in alkali Bose-Einstein
condensates \cite{BEC}. 
This is possible thanks to the remarkable degree of control
in optical manipulations of the alkali condensates and due to the fact 
that their thermal components can be made very small \cite{Jin}.
Slow light \cite{Hau} plays the role of a single excitation 
of the normal component. The force due to the Aharonov-Bohm
effect of slow light we call the optical Iordanskii force. 

\section{Proposed Experiment}

\begin{figure}
\includegraphics[width=19.5pc]{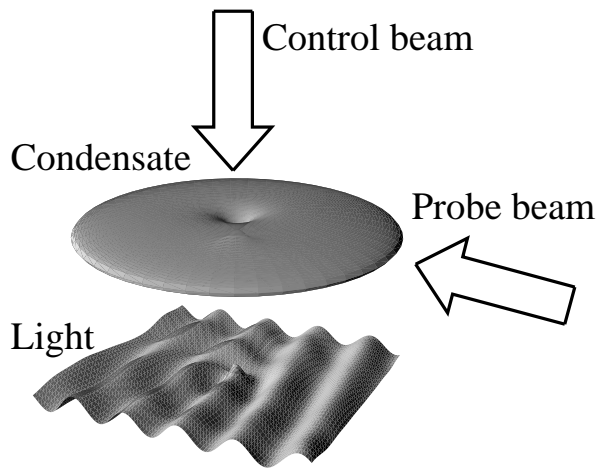}
\caption{\label{fig1} Diagram of the proposed experiment. The
control beam slows down the probe light that experiences an
optical Aharonov-Bohm effect caused by a vortex in a Bose-Einstein
condensate. We investigate the dynamics of the vortex.}
\end{figure}

The envisioned experiment is based on
Electromagnetically Induced Transparency (EIT) \cite{EIT,Leo}.
Suppose that the condensate, carrying a vortex \cite{Vortices},
is illuminated by a uniform control beam in the direction of the
vortex line, see Fig.\ 1. The control light is in resonance with two excited
states of each of the condensate atoms. The beam controls the
group velocity $v_g$ of a second beam, the probe beam, that
propagates perpendicularly to the vortex line. 
The energy of the probe beam is transferred into the control light.
The momentum transfer of the probe does not remove atoms from
the condensate, as long as $v_g$ is large compared 
with the recoil velocity of the atoms \cite{Dutton}.
The probe should be
a stationary wave with a frequency $\omega$
in the laboratory frame that
exactly matches the atomic transition frequency $\omega_0$ 
between the
ground state and one of the excited states coupled by the control
beam. In the flowing condensate the probe light is subject to the
Doppler effect \cite{Leo}. As long as the vortex flow ${\bf u}$
is slow compared with $v_g$, the probe experiences an optical
Aharonov-Bohm effect caused by the local Doppler detuning
\cite{LP} and forms a distinct interference pattern \cite{AB}.
One can monitor {\it in situ} the position of the vortex core by
phase-contrast microscopy \cite{Andrews} of the control and the
probe beam, providing two independent measurements. 
Let us calculate the forces on the vortex.

\section{Light and matter}

First we develop a phenomenological  Lagrangian theory 
of the interaction of slow light with Bose-Einstein condensates. 
This theory is independent of the microscopic mechanisms 
used to slow down light, experimental techniques that may 
change in the future. Moreover, with minor modifications, 
the theory is applicable to other wave phenomena as well.
We describe the probe light by a real scalar field
$\varphi$ of fixed polarization, where $\varphi$ represents
the electric field in units of the vacuum noise,
$E_p=(\hbar/\varepsilon_0)^{1/2}\omega_0\varphi$,
with $\omega_0$ being the resonance frequency of EIT. 
Consider a slow-light medium at rest without taking into account 
the forces of light on the atoms.
According to Ref.\ \cite{Leo}, slow light 
obeys the principle of least action \cite{LL2}
with the Lagrangian density
\begin{equation}
\label{ll}
{\mathscr L}_L=\frac{\hbar}{2}\,
\Big((1+\alpha)(\partial_t\varphi)^2-c^2(\nabla\varphi
)^2-\alpha\,\omega_0^2\varphi^2 \Big) \,,
\end{equation}
because the corresponding  Euler-Langrange equation \cite{LL2}
does indeed describe the propagation of the probe field \cite{Leo}.
The Lagrangian density (\ref{ll}) depends on the 
group index $\alpha$ that is related to the
group velocity of the probe light \cite{Leo}
\begin{equation}
v_g=\frac{c}{1+\alpha}\,.
\end{equation}
The group index $\alpha$ is proportional to the number of atoms per unit
volume, $\rho$, 
and $\alpha$ is inversely proportional to the intensity of the
control beam \cite{Leo}.
The two light beams induce atomic dipoles with 
density \cite{Leo}
\begin{equation}
\rho_{_D}=
\alpha\,\frac{\varepsilon_0|E_p|^2}{\hbar\omega_0}=\frac{1}{2}
\left|\frac{\Omega_p}{\Omega_c}\right|^2 \rho \,.
\end{equation}
Here we have calibrated the intensities of the probe and the control
fields in terms of their Rabi frequencies \cite{Leo} $\Omega_p$ and
$\Omega_c$, respectively. The control beam should dominate and, in
practice \cite{Hau}, $|\,\Omega_p /\Omega_c|^2$ does not exceed
$10^{-1}$. In this case absorption is sufficiently suppressed. 

Consider now a moving slow-light medium with velocity profile ${\bf u}$. 
To find the effective Lagrangian density of slow light,
${\mathscr L}_{\mbox{\small\it eff}}$, we assume that 
the Lagrangian  (\ref{ll}) is valid in locally co-moving frames and
we perform local Lorentz transformations to the laboratory frame.
We note that $(\partial_t\varphi)^2-c^2(\nabla\varphi)^2$
is a Lorentz scalar and we replace the remaining term 
$\alpha(\partial_t\varphi)^2$
by $\alpha[(\partial_t + {\bf u}\cdot \nabla)\varphi]^2$.
Consequently, we get, to first order in $u/c$, in the laboratory frame 
\begin{equation}
\label{leff}
{\mathscr L}_{\mbox{\small\it eff}} = {\mathscr L}_{L}-\alpha{\bf
P \cdot u}\,,
\end{equation}
expressed in terms of the vector
\begin{equation}
{\bf P}=-\hbar (\partial_t \varphi)\nabla\varphi \,.
\end{equation}
As shown in Ref.\ \cite{Leo}, 
${\bf P}$ describes the energy flux of the 
probe light, the Poynting vector. 
The resulting Euler-Lagrange equation 
governs the propagation of slow light in moving media
with a given velocity profile ${\bf u}$.

A Bose-Einstein condensate is a delicate quantum fluid that 
is easily influenced by optical forces. 
Here we cannot assume that the condensate represents a medium
with given dielectric properties and with given velocity profile.
Instead of regarding ${\bf u}$ as fixed we must consider light 
and matter as a combined dynamical system,
\begin{equation}
{\mathscr L}={\mathscr L}_{L}+{\mathscr L}_{M}\label{la},
\end{equation}
with the Gross-Pitaevskii Lagrangian \cite{BEC}
\begin{equation}
{\mathscr L}_{M} = -\rho\left(\hbar \dot S + \frac{m}{2}u^2 +
\frac{\hbar^2}{2m}(\nabla \sqrt{\rho})^2 + \frac{g}{2}\,\rho +
V\right). \label{gp}
\end{equation}
Here $S$ and $\rho$ describe the phase and the density profiles 
of the condensate, respectively, 
$m$ is the atomic mass, $g$ characterizes the atom-atom
collision strength and $V$ denotes an external potential.
For each atom, the interaction potential with the probe light
is proportional to the induced atomic dipole moment. 
In EIT this dipole moment vanishes when the probe light
is exactly on resonance \cite{Leo}. 
The detuning from the resonance 
frequency $\omega_0$ induces an optical force. 
Since, within the spectral window of EIT, 
the dipole moment depends linearly on the detuning \cite{Leo},
the potential is linear in the Doppler detuning
caused by the moving atom.
Therefore, the interaction potential of each atom 
is proportional to the scalar product 
between the Poynting vector and the atom's velocity.
Consequently \cite{LL2},
slow light and matter interact with each other as if
the Poynting vector were a vector potential,
similar to the R{\"o}ntgen interaction of 
electromagnetric fields in Bose-Einstein condensates 
\cite{LProentgen}. 
An interaction of vector-potential nature appears
in the relation between the flow ${\bf u}$ 
and the phase $S$ of the wavefunction \cite{LL3}
\begin{equation}
m {\bf u}=\hbar \nabla S+\alpha_0 {\bf P}
\label{coupling}
\end{equation}
with the coupling constant $\alpha_0$. 
We could determine $\alpha_0$ from the specific theory of EIT \cite{Leo}.
More elegantly, we chose $\alpha_0$ such that the 
propagation of slow light resulting from the Lagrangian (\ref{la})
agrees with the dynamics of the effective
Lagrangian (\ref{leff}) of slow light in moving media.
First we note that ${\bf u}$ does indeed describe a flow, 
\begin{equation}
\partial_t \rho+\nabla(\rho {\bf u})=0 \,,\label{continuity}
\end{equation}
according to the Euler-Lagrange equation with respect to $S$.
Then we find that the variations of ${\mathscr L}_{M}$ and ${\mathscr
L}_{\mbox{\small\it eff}}$ with respect to ${\bf P}$ agree, if 
\begin{equation}
\alpha = \alpha_0\rho\,.
\end{equation}
Consequently, the Lagrangian (\ref{la}) generates the wave
equation of slow light in a moving medium with flow ${\bf u}$
\cite{Leo} which justifies the postulated coupling (\ref{coupling})
between light and matter.

\section{Aharonov-Bohm interaction} 

In the original setup of
Aharonov and Bohm \cite{AB} a magnetic vortex acts on a matter
wave. Here we study a situation close to the opposite case where
a material vortex acts on an electromagnetic wave. Throughout the
rest of the paper we regard both the condensate and the slow
light as homogeneous in the direction ${\bf e}$ of the vortex
line, assuming an effective two-dimensional model with ${\bf
z}=(x,y)$ and regarding all densities as effectively two-dimensional. 
Consider first a fixed vortex \cite{Vortextheory} with flow
\begin{equation}
{\bf u}_0=\frac{h}{m} \frac{{\bf e}\times {\bf z}}{|{\bf z}|^2}
\,.
\end{equation}
Outside of the vortex core the density $\rho_0$ approaches the
bulk density $\rho_{_B}$ and inside the core the density drops to
zero \cite{Vortextheory}. Consider monochromatic slow light with
frequency $\omega_0$ in the laboratory frame. In the most
elementary model for the optical Aharonov-Bohm effect
\cite{LeoPiw} the positive frequency part $\varphi^{(+)}$
represents a plane wave with a phase slip,
\begin{equation}
\varphi_0^{(+)}=\sqrt{\frac{\varepsilon_0}{2\hbar}}\,
\frac{E_0}{\omega_0}\, \exp \Big(i{\bf k\cdot z}-i\nu \arg{\bf
z}-i\omega_0 t\Big)
\end{equation}
characterized by the effective Aharonov-Bohm flux quantum
\cite{LP}
\begin{equation}
\nu=\alpha\,\frac{\hbar\omega_0}{mc^2}\,.
\end{equation}
The light should be slow enough to experience a noticeable effect
but $v_g$ should exceed the flow speed in the bulk (about
$10^{-3}$ m/s) by at least two orders of magnitude, in order to
avoid a Doppler-detuning out of the narrow frequency range of EIT
\cite{Leo}. In practice \cite{Hau} the maximal $\nu$ can be about
$10^{-1}$.

Consider a vortex moving relative to the bulk condensate with core
position ${\bf z}_0$ and velocity ${\bf v}=d{\bf z}_0/dt$. We
assume that $v$ is much smaller than the speed of sound such that
the vortex can adjust itself to move coherently,
\begin{equation}
\rho=\rho_0 ({\bf z}-{\bf z}_0)
\end{equation}
and
\begin{equation}
{\bf u}={\bf u}_0({\bf z}-{\bf z}_0) \quad\mbox{outside of the
core.}
\label{outside}
\end{equation}
Simultaneously, the light follows the wandering vortex,
\begin{equation}
\varphi_0^{(+)}=\varphi_0^{(+)}({\bf z}-{\bf z}_0)\,.
\end{equation}
The quantum fluid must continuously refill the core of the moving
vortex in order to satisfy the continuity equation
(\ref{continuity}) when $\rho$ varies,
\begin{eqnarray}
0=\partial_t\rho+\nabla(\rho{\bf u})&=&
({\bf u}-{\bf v})\cdot\nabla\rho + \rho\nabla\cdot{\bf u}
\nonumber\\
&=& ({\bf u}-{\bf v})\cdot\nabla\rho\,.
\end{eqnarray}
Therefore, the velocity profile (\ref{outside}) cannot be correct 
inside of the vortex' core. We see that ${\bf u}$
approaches ${\bf u}_0 ({\bf z}-{\bf z}_0)+{\bf v}$ in the core.
To account for the condensate's back flow 
in the Gross-Pitaevskii Lagrangian (\ref{gp})
we split the kinetic energy density $\rho{m}u^2/2$ into the two parts
$\rho_{_B}{m}u^2/2$ and $(\rho-\rho_{_B}){m}u^2/2$.
The first component describes the bulk of the condensate
where we assume that the velocity profile (\ref{outside}) is valid.
The second component describes the back flow.
Here we replace $(\rho-\rho_{_B}){m}u^2/2$ by
$(\rho-\rho_{_B}){m}({\bf u}+{\bf v})^2/2$.
In this way we obtain from
the Gross-Pitaevskii Lagrangian (\ref{gp})
\begin{eqnarray}
{\mathscr L}_M&=& 
\rho \hbar{\bf v}\cdot\nabla S
 -\frac{m}{2}(\rho_0-\rho_{_B})({\bf u} + {\bf v})^2
-  \frac{m}{2} \rho_{_B}u_0^2
\nonumber\\
&& - \rho \Big( \frac{\hbar^2}{2m} (\nabla
\sqrt{\rho})^2 + \frac{g}{2}\,\rho + V \Big)
\,,
\label{varan}
\end{eqnarray}
where $S$ describes the phase profile of the vortex 
(\ref{outside}) according to the relation (\ref{coupling}).
The variational ansatz (\ref{varan})
reduces the principle of least action for
the light and matter waves to the mechanics of a point particle,
the vortex core, minimizing the action $\int L\, dt$ with the
Lagrangian
\begin{equation}
L = \int\int{\mathscr L}\, dx\, dy 
= \frac{m_v}{2} v^2 + {\bf
v}\cdot {\bf A}_v - U
\label{lvortex}
\end{equation}
and
\begin{eqnarray}
m_v &=& m\int\int(\rho-\rho_{_B})\,dx\,dy \,,\nonumber\\
{\bf A}_v &=& \int\int(m\rho_{_B} {\bf u}-\alpha {\bf P})\,dx\,dy 
\,,\nonumber\\
U &=& U_0 - \hbar c^2 \int\int\left|\nabla\varphi^{(+)}\right|^2
dx\,dy\,.
\end{eqnarray}
The vortex mass $m_v$ accounts for the missing mass of the vortex
core (a negative quantity) and it agrees exactly with Duan's
vortex mass in superfluids \cite{Duan}. The mass depends
logarithmically on the size of the condensate and is best inferred from 
experimental observations. 
In fact, our setup can be used to measure the vortex mass.
The vortex moves in effective vector and scalar potentials 
${\bf A}_v$ and $U$. In $U$ we have singled out the kinetic energy 
of the light wave, $\hbar c^2 (\nabla\varphi)^2/2$, 
and we have averaged over the rapid optical oscillations. 
$U_0$ accounts for the remaining terms of the Lagrangian.

\section{The forces}

We use the point-particle Lagrangian
(\ref{lvortex}) to find the forces acting on the moving vortex.
The forces contain the curl and the gradient of the potentials
${\bf A}_v$ and $U$ with respect to the vortex coordinate ${\bf
z}_0$. Only the long-ranging vortex terms in ${\bf A}_v$ and $U$
will depend on ${\bf z}_0$. We calculate the effective
magnetic-type field
\begin{eqnarray}
\nabla_0\times{\bf A}_v &=& -\int\int\nabla\times(m\rho_{_B} {\bf
u}-\alpha {\bf P})\,dx\,dy \nonumber\\
&=& -\hbar {\bf e}\, (\rho_{_B}+\nu \rho_{_D})\oint \frac{{\bf
e}\times({\bf z}-{\bf z}_0)}
{|{\bf z}-{\bf z}_0|^2}\cdot d {\bf z} \nonumber\\
&=& -h {\bf e}\, (\rho_{_B}+\nu \rho_{_D})
\end{eqnarray}
and the effective electric-type field, using the abbreviation
${\bf v}_g= v_g {\bf k}/(\omega_0/c)$,
\begin{eqnarray}
-\nabla_0 U &=& \hbar\nabla_0\left({\bf v}_g\cdot \int\int
\rho_{_D}\nu\, \frac{{\bf e}\times({\bf z}-{\bf z}_0)} {|{\bf
z}-{\bf z}_0|^2}\,dx\,dy \right)\nonumber\\
&=& \hbar {\bf v}_g \times \left(\nabla_0 \times \int\int
\rho_{_D}\nu\, \frac{{\bf e}\times({\bf z}-{\bf z}_0)} {|{\bf
z}-{\bf z}_0|^2}\,dx\,dy\right)\nonumber\\
&=& -\hbar \nu\rho_{_D} {\bf v}_g \times {\bf e} \left(\oint
\frac{{\bf e} \times ({\bf z}-{\bf z}_0)}{|{\bf z}-{\bf z}_0|^2}
\cdot d{\bf z}\right)
\nonumber\\
&=& h \nu \rho_{_D} {\bf e} \times {\bf v}_g \,.
\end{eqnarray}
Consequently,
\begin{equation}
m_v\frac{d{\bf v}}{dt} = h {\bf e} \times \Big((\rho_{_B} +
\rho_\nu){\bf v} + \rho_\nu{\bf v}_g\Big). \label{dyn}
\end{equation}
with 
\begin{equation}
\rho_\nu = \nu \rho_{_D}= 
\frac{\varepsilon_0|E_0|^2}{mv_g^2}\quad 
\mbox{for}\,\, v_g \ll c\,.
\end{equation}
Two forces act on the vortex --- the total Magnus force $ h{\bf
e}\times(\rho_{_B}+\rho_\nu){\bf v}$ and the optical Iordanskii
force \cite{Stone2} $h\rho_\nu {\bf e}\times{\bf v}_g$. In the
sonic analog \cite{Stone2} of the Aharonov-Bohm effect in
superfluids, $\rho_\nu$ and ${\bf v}_g$ correspond to the
density $\rho_n$ and the flow of the normal component, 
because here $\rho_n$ is the energy density divided by $mc_s^2$
where $c_s$ denotes the speed of sound, assuming that the normal
component consists of elementary excitations \cite{BEC}.
In this case, however, the non-dissipative dynamics (\ref{dyn}) 
has never been observed experimentally. 
An alkali Bose-Einstein condensate can be made such that it hardly
possesses a noticeable normal component \cite{Jin}, thus allowing
friction-less vortex motion. In fact, slow light 
assumes the role of the normal component.
The key advantage of light is the remarkable 
degree to which it can be controlled and observed.

As a result of the optical Aharonov-Bohm interaction the vortex
behaves like a point particle in homogeneous magnetic and
electric fields \cite{LL2}. We write down the solution \cite{LL2}
of the dynamics (\ref{dyn}) in the complex notation $z=x+iy$,
defining our system of coordinates such that the slow light is
incident in negative $x$ direction,
\begin{equation}
z_0(t) = \frac{v_{_D}}{i\omega_c}\, (e^{-i\omega_c t} - 1) +v_{_D}
t\,.
\end{equation}
The vortex orbits with the cyclotron frequency
\begin{equation}
\omega_c = \frac{h}{|m_v|}(\rho_{_B} + \rho_\nu)=\frac{h\rho_{_B}}{|m_v|}
\end{equation}
and drifts against the direction of the incident slow light with the velocity
\begin{equation}
v_{_D} = \frac{\rho_\nu}{\rho_{_B} + \rho_\nu} v_g =
\frac{1}{2} \left| \frac{\Omega_p} {\Omega_c} \right|^2 v_{_R}
\,,\quad v_{_R} = \frac{\hbar\omega_0}{mc} \,,
\end{equation}
neglecting $\rho_\nu$ in the Magnus force. According to the
numerical simulations reported in the Appendix the cyclotron
frequency is in the order of $2\pi\times 10{\rm Hz}$ for a sodium 
condensate of a few $10^{13} {\rm m}^{-2}$ 2D density.
The drift velocity is proportional to the atomic recoil velocity 
$v_{_R}$ (about $3.5\times 10^{-2}{\rm m}/{\rm s}$ for sodium) 
and depends on the ratio of the probe and control intensities 
given in terms of the Rabi frequencies \cite{EIT}.  
The initial acceleration
due to the optical Iordanskii force, $v_{_D}\omega_c$, 
is in the order of
$10^{-2}{\rm m}/{\rm s}^2$ 
for  $|\Omega_p/\Omega_c|^2 = 10^{-2}$,
which is substantial for a condensate of $10^{-4}{\rm m}$ size.
In this case the radius of the cyclotron motion, 
$v_{_D}/\omega_c$, reaches several $10^{-6}{\rm m}$.
Our numerical simulations support our analytical 
results and show that the vortex remains intact, see the Appendix.
Therefore, the optical Aharonov-Bohm effect \cite{LP} of slow
light \cite{Hau} in alkali Bose-Einstein condensates offers
excellent prospects for seeing a Iordanskii-type force in
real-time action.

\section{Summary}

A vortex in an alkali Bose-Einstein condensate
generates the optical analog \cite{LP} of the Aharonov-Bohm
effect \cite{AB} when illuminated with slow light \cite{Hau}. Far
away from the vortex' core, light rays travel along straight
lines, but a slow-light wave develops a distinctive interference
pattern \cite{LeoPiw}. Despite the apparent absence of a
kinetic force on the light rays, the vortex will react and is
accelerated, a behaviour in sharp contrast to the magnetic
Aharonov-Bohm effect \cite{AB} where no forces are exchanged. A
sonic Aharonov-Bohm effect \cite{SAB} leads to a back reaction as
well, called the Iordanskii force \cite{Stone2}, that is believed
to lie at the heart of the mutual friction \cite{Iordanskii}
between vortices and the normal component in liquid helium
\cite{Friction}, although the force has been subject to bitter
controversy \cite{Sonin,Controversy}. In contrast to the sonic
Iordanskii force \cite{Stone2}, the reaction of a vortex on slow
light can be observed directly as a mechanical force, within
realistic experimental parameters. Such an experiment would
demonstrate convincingly a fundamental difference between the
magnetic Aharonov-Bohm effect \cite{AB} and its analogs
\cite{SAB,OAB,Other}, despite many similarities.

\section*{ACKNOWLEDGEMENTS}

We thank O. B\"uhler, S. Lee,  L. Santos, A. Shelankov, M. Stone, 
and G. E. Volovik for discussions. 
The paper was supported by the Leverhulme Trust, 
by EPSRC, by the Royal Society of Edinburgh, and by the
ESF programme Cosmology in the Laboratory.

\section*{APPENDIX}

We performed numerical simulations to test our analytical results
and to determine the effective vortex mass.
We solved the equations of motion that result from the 
complete Lagrangian density (\ref{la}),
the 2D Gross-Pitaevskii equation for the condensate $\psi$ 
coupled to the slow-light field 
\begin{equation}
i\hbar\frac{\partial \psi}{\partial t} = \frac{1}{2m}\left(-i\hbar\nabla + \alpha_0{\bf P}\right)^2\psi + V\psi+ g|\psi|^2\psi
\end{equation}
and the propagation equation of slow light, 
\begin{eqnarray}
\Big(\partial_t(1+\alpha)\partial_t &-& c^2
\nabla^2 + \alpha\,\omega_0^2 
\nonumber\\
&+&
\partial_t\alpha\, {\bf u}\cdot\nabla 
+ \nabla\cdot \alpha {\bf u}\,\partial_t \Big)
\varphi = 0\,.
\end{eqnarray}
The condensate was hold in a hard-wall cylindric trap, to have a
homogeneous density, in order to compare the numerical results
with our analytic theory. We also simulated traps 
with harmonic-oscillator potential $V$ and got similar results.
We generated an exact Gross-Pitaevskii vortex as the initial condition for $\psi$.
The light was modeled  as a long smooth pulse that enters the condensate from the right and that propagates to the left. In our simulations we have been only able to cover the initial stage of the dynamics, because of the vast difference in the scales involved (condensate size versus pulse length).  The simulations reach the limits of top-performance single-processor PCs. (To simulate a few milliseconds takes a few hours.) Figure 2 shows a comparison of the vortex before and after our simulation period, showing clearly that the vortex has remain intact. The sub-figure below compares the position of the vortex' core (determined as the minimum of the density for snapshots taken at discrete time steps) with our simple analytical prediction. The only fitting parameters used are the vortex mass ($1.0\times 10^5$ times the atomic mass of sodium, $3.8\times 10^{-26}{\rm kg}$) and the buildup time for the optical Aharonov-Bohm effect that defines the time zero in our analytical fit (we find $3.6\times 10^{-3}{\rm s}$). The vortex mass is not well defined in our simple model. (The  mass diverges logarithmically with the size of the condensate.) In fact, our proposed experiment can be applied to measure the vortex mass. The initial light pulse causes the condensate to oscillate with the trap frequency. Averaged over the trap oscillations the agreement between our simple analytical result and the numerics is good.

\begin{figure}
\includegraphics[width=20.5pc]{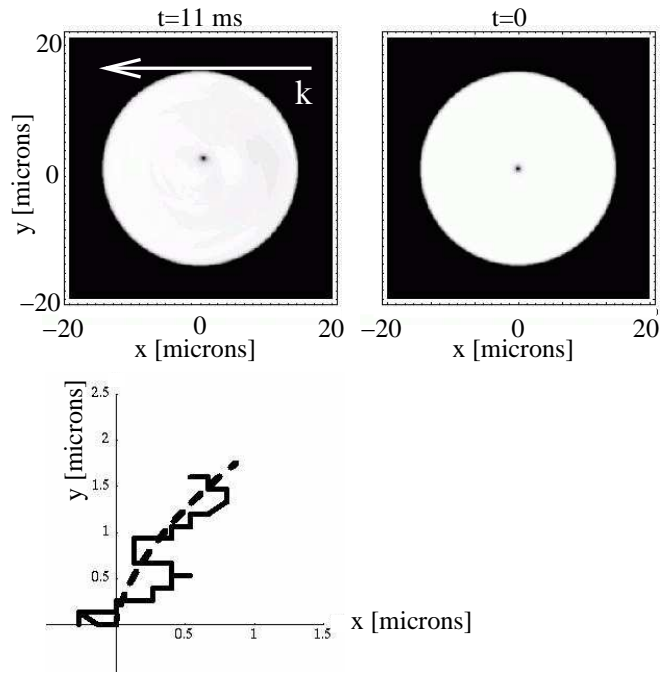}
\caption{\label{fig2} 
Simulations of the initial stage of the vortex dynamics.
The figure illustrates the combined effect of the 
optical Iordanskii force and the Magnus force.
We simulated the complete dynamics of the slow light
and the condensate, as described in Eqs.\ (29) and (30).
The upper pair of sub-figures shows a snapshot 
of the evolution of the vortex in a cylindric condensate.
Regions of low densities, such as the vortex' core, appear black.
The lower sub-figure compares the simulated position of the core
(line) with our analytical prediction (dashed line)
as a function of time. }
\end{figure}

Our numerics indicates that the vortex remains intact. 
The absorption in EIT is small. The generated vortex dynamics is substantial and measurable in alkali Bose-Einstein condensates. 
Therefore we have all reasons to believe that the proposed experiment is feasible.


\end{document}